\documentclass[a4paper]{jpconf}
\usepackage{graphicx}
\newcommand{\newc}{\newcommand}
\newc{\ra}{\rightarrow}
\newc{\lra}{\leftrightarrow}
\newc{\be}{\begin{equation}}
\newc{\ee}{\end{equation}}
\newc{\ba}{\begin{eqnarray}}
\newc{\ea}{\end{eqnarray}}
\newc{\nn}{\nonumber}
\begin{document}
\title{The origin of discrete symmetries in F-theory models}

\author{ George K. Leontaris}

\address{Theoretical Physics Division, Ioannina University, GR-45110 Ioannina, Greece}

\ead{leonta@uoi.gr}

\begin{abstract}
While non-abelian  groups are undoubtedly the cornerstone of Grand Unified Theories (GUTs),
phenomenology shows  that the role of abelian and discrete symmetries is equally important in
model building. The latter are the appropriate tool to suppress undesired proton decay operators
and various flavour violating interactions, to generate a hierarchical fermion mass spectrum, etc.
In F-theory,  GUT symmetries are linked to the singularities of the elliptically fibred K3 manifolds;
they are of ADE type and have been extensively discussed in recent literature.  In this context,
abelian and discrete symmetries usually arise either as a subgroup of the non-abelian symmetry or from a
non-trivial Mordell-Weil group associated to rational sections of the elliptic fibration. In this note
 we give a short overview of the current status and focus in models with rank-one Mordell-Weil group.
\end{abstract}

\section{Introduction}

Discrete symmetries play a vital role in model
building~\cite{Krauss:1988zc,Ibanez:1991pr,Nath:2006ut,Lee:2011dya}.  Over the past few decades
they have been widely used to restrict the superpotential and suppress the exotic
interactions of numerous proposed effective theories.
In the Standard Model, as well as in old GUTs,
abelian factors and discrete symmetries of $Z_N$ type where imposed to forbid
dangerous Lepton and Baryon number violating operators. In more recent scenarios,
non-abelian discrete groups where introduced to interpret the mixing  properties of
the neutrino sector~\cite{Ishimori:2010au,Altarelli:2010gt,King:2013eh}.
In the field theory context, these symmetries were postulated
purely on phenomenological grounds.  However, it is not clear whether such global
(including discrete) symmetries can exist~\cite{Banks:2010zn}.
In this respect, it would be interesting to
investigate whether such symmetries can be justified in the context of string theory.
Recently, considerable work in this direction has been
devoted~\cite{Ibanez:2012wg}-\cite{Mayrhofer:2014laa}. A fascinating possibility
in particular arises in F-theory constructions where symmetries are tightly
connected to the  elliptically  fibred  internal  space. This compact space is
a four-dimensional complex manifold
 (a fourfold) while the gauge symmetries  are linked to its singularities. Hence,
 we may consider that all symmetries, including the discrete part, are associated
 to the geometric properties of the fourfold.
In the present talk I will discuss the origin of discrete symmetries in F-models.
I will start with a short description  of the  basic features of F-model building focusing in
particular to the properties  related to the  elliptic curves.

\section{Elliptic Curves}

F-theory~\cite{Vafa:1996xn}  is an exciting reformulation of String Theory in a 12-dimensional
space which consists of the 4 space-time dimensions and an 8-dimensional internal elliptically
fibred compact space. Because of its relation to the theory of elliptic curves and in
particular their complex representations, effective F-theory models are endowed
with many interesting properties. In the present talk I will mainly focus on
some features of the abelian and discrete
symmetries that emerge from the rational sections of the elliptic curves.

Many of the properties that will be discussed are related to
 rational points on curves. A point is said to be rational  if its coordinates are rational,
while a rational curve is defined by an equation with rational coefficients.
It is trivial  to find the rational points on lines and conics. Consider for example
the equation of the unit circle $x^2+y^2=1$. Choosing a rational point on it,
-let it be $(-1,0)$-  we can draw a line which intersects the circle at $(x,y)$.
We can map this pair on the line identified with the vertical axis $y$ by the transformation
\[ x=\frac{1-t^2}{1+t^2},\;  y =\frac{2t}{1+t^2}, \]
\begin{figure}[h]
\centering
\includegraphics[scale=.8]{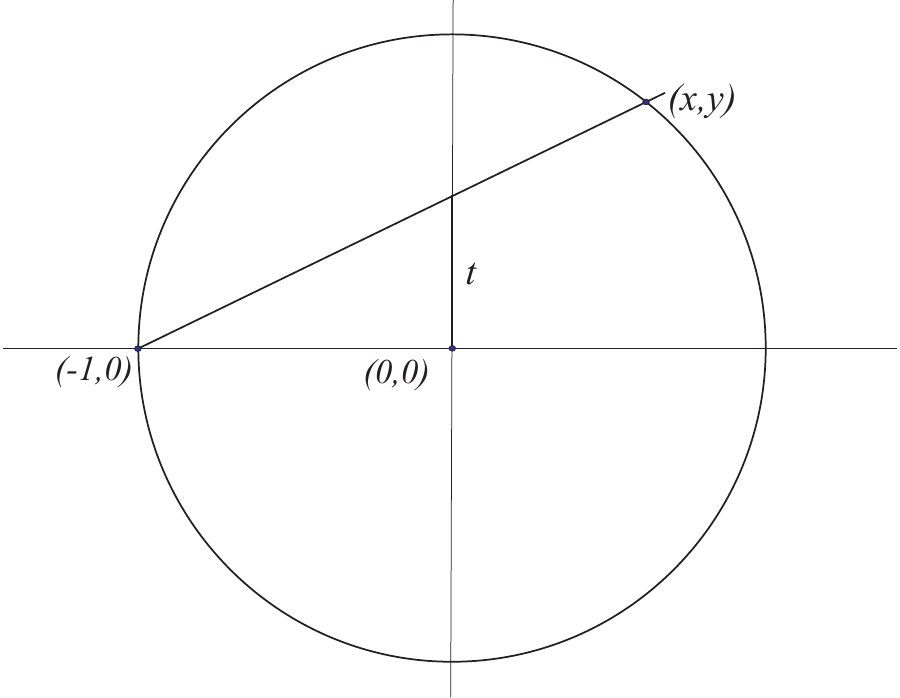}
\caption{Rational points on conics.}
\label{Conic}
\end{figure}
This way,  all rational  points $(x,y)$ on the curve can be determined
in terms of the rational values of the parameter $t$.
Because of this  correspondence between the rational points,  we say that
the curve (in this case the conic) is {\bf birationally equivalent} to the line.  A rational curve is
also called a curve of genus zero. Every genus zero curve is birationally equivalent either to a
conic or to a line.

\begin{figure}[h]
\centering
\includegraphics[scale=.4]{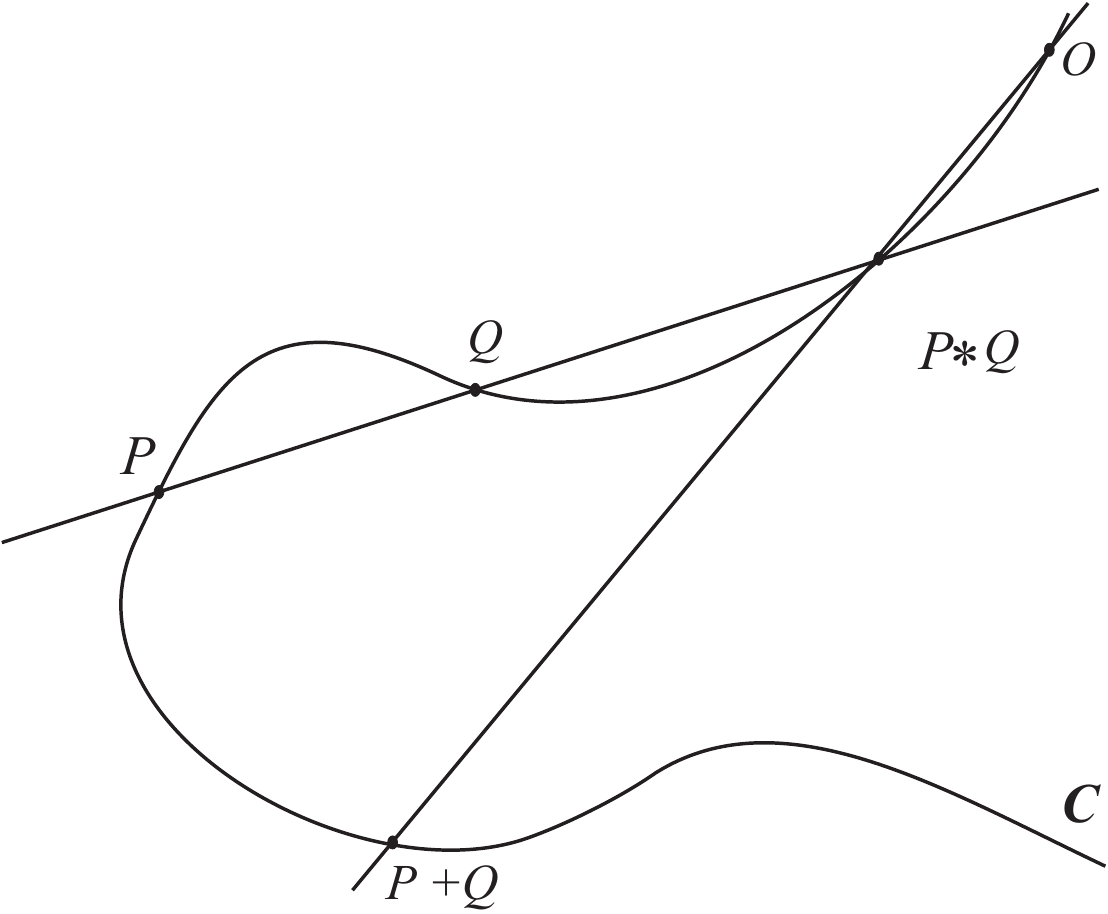}\;\;\;\;\;\;\;\;\;\;\;\;\;\;\;\;\;\;\;\;\;\;\;\;
\includegraphics[scale=.4]{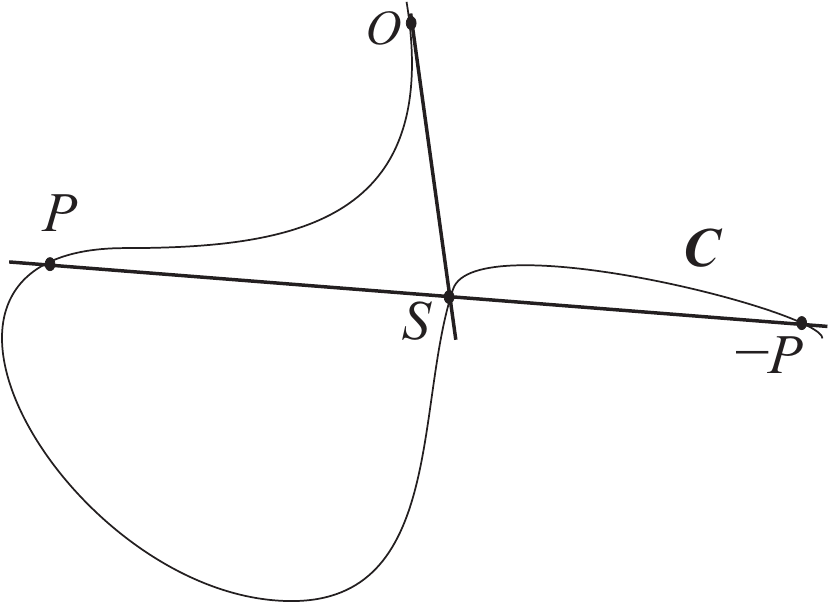}
\caption{The group structure of rational points on elliptic curves. The law of addition.}
\label{GroupLaw}
\end{figure}

We proceed now to the  elliptic curves which are
  described by a cubic equation  whose most general form can be written as
\be
{\cal C}:\;\;\; \sum_{n=0}^3\sum_{m=0}^n a_{m,n} x^n y^{n-m} =0\label{GELC}
\ee
The identification of the rational points  on a given elliptic curve ${\cal C}$ of general type
is much more complicated compared to the conic.
We know however, that the rational points of ${\cal C}$  exhibit a group structure.
According to Mordell's theorem,\\
  {\it  If a non-singular elliptic curve ${\cal C}$ has a  rational point then the group of rational points
can be finitely generated.}\\
In other words, there is a finite number of elements generating the whole group.
The group structure is depicted here in figure~\ref{GroupLaw} which can be defined as follows:
Let $P,Q$ two rational points on ${\cal C}$.  Drawing  the
line joining these two points,  we can find another one  at the third intersection of
the line with the curve ${\cal C}$.  I designate this point with $P*Q$.
Suppose now we are given a rational point $O$ on ${\cal C}$ that we can identify this to be the zero element
of the group. The line from $O$ to $P*Q$  intersects ${\cal C}$ on another point which,
as can be proved~\footnote{See for example standard textbooks such as~\cite{TateSilv,Silverman,Cassels}.},
 under the group law is the point  $P+Q$.
To find the opposite element with respect to the addition law of the group, we draw a tangent
 to the zeroth element $O$ which intersects
${\cal C}$ at some point called here  $S$. One  can prove that the opposite to $P$  is identified with the third
 intersection of the $PS$-line and curve ${\cal C}$, so that  $ P+(-P)=O$.

The general form (\ref{GELC}) of the elliptic equation is rather too complicated. Fortunately it can be shown that
any  cubic equation with a rational point on it can be brought to the
 Weierstra\ss \,  form
\be y^2=x^3+f x +g \label{WeiF}
\ee
We can readily check that the Weierstra\ss\, form is symmetric with respect to the $x$-axis.
Moreover, the zeroth element of the group can be taken to infinity while the sum of two points
is just the  reflection (w.r.t. $x$-axis) of the third intersection point of the line PQ with ${\cal C}$.
There are two important quantities characterising the elliptic curves. These are:
\\
$\bullet$  The discriminant
\ba
{ \Delta} &=& 4\,{ f}^3+27\, { g}^2\label{Discr}
\ea
which classifies the singularities on the matter curve. In particular,
when $\Delta\ne 0$  curves are non-singular and may have one or three real roots.
When $\Delta\,=\,0$  curves are singular. Singularities are of nodal or cuspidal type.
\\
$\bullet$
The j-invariant (modular invariant) function
\be
j(\tau)=\frac{4(24 f)^3}{4f^3+27 g^2}\; = \; \frac{4(24 f)^3}{\Delta}  \label{Jinv}
\ee
which takes the same value for equivalent elliptic curves
characterised by the $SL(2,Z)$ transformations  $ \tau \to \frac{a\tau+b}{c\tau+d}$.

What happens when we consider complex coefficients (functions) $f, g \in \bf C$? In this case, it can proved that a
complex elliptic curve ${\cal C}$ is a genus 1 closed surface with a marked point on it corresponding to its
neutral element (point to infinity). Defining a modulus $\tau$ as usually, the torus, hence ${\cal C}$,
 is equivalent  to the lattice  $(1,\tau)$.  With respect to the previous analysis of elliptic curves, we
 distinguish two cases:
 The complex analogue of a real elliptic curve with  non-singular points is  a torus without singularities.
 On the contrary, if the real elliptic curve has singular points then its complex
 equivalent is a torus with a pinched radius.


\section{F-theory and Elliptic Fibration}

In this section,  a few  basic features of F-theory~\cite{Vafa:1996xn} are described
 which are useful to the subsequent analysis. The shortest (however incomplete) definition of
 F-theory is that it is the geometrisation of the type II-B superstring.  The  type IIB string
 is distinguished  by its closed string spectrum (which differs, say, from II-A case). It is obtained by combining
  L- and R- moving open strings with the truncated two types of boundary conditions, namely the
  Neveu-Schwarz  (antiperiodic) and Ramond  (periodic) boundary conditions.

\begin{figure}[h]
\centering
\includegraphics[scale=.65]{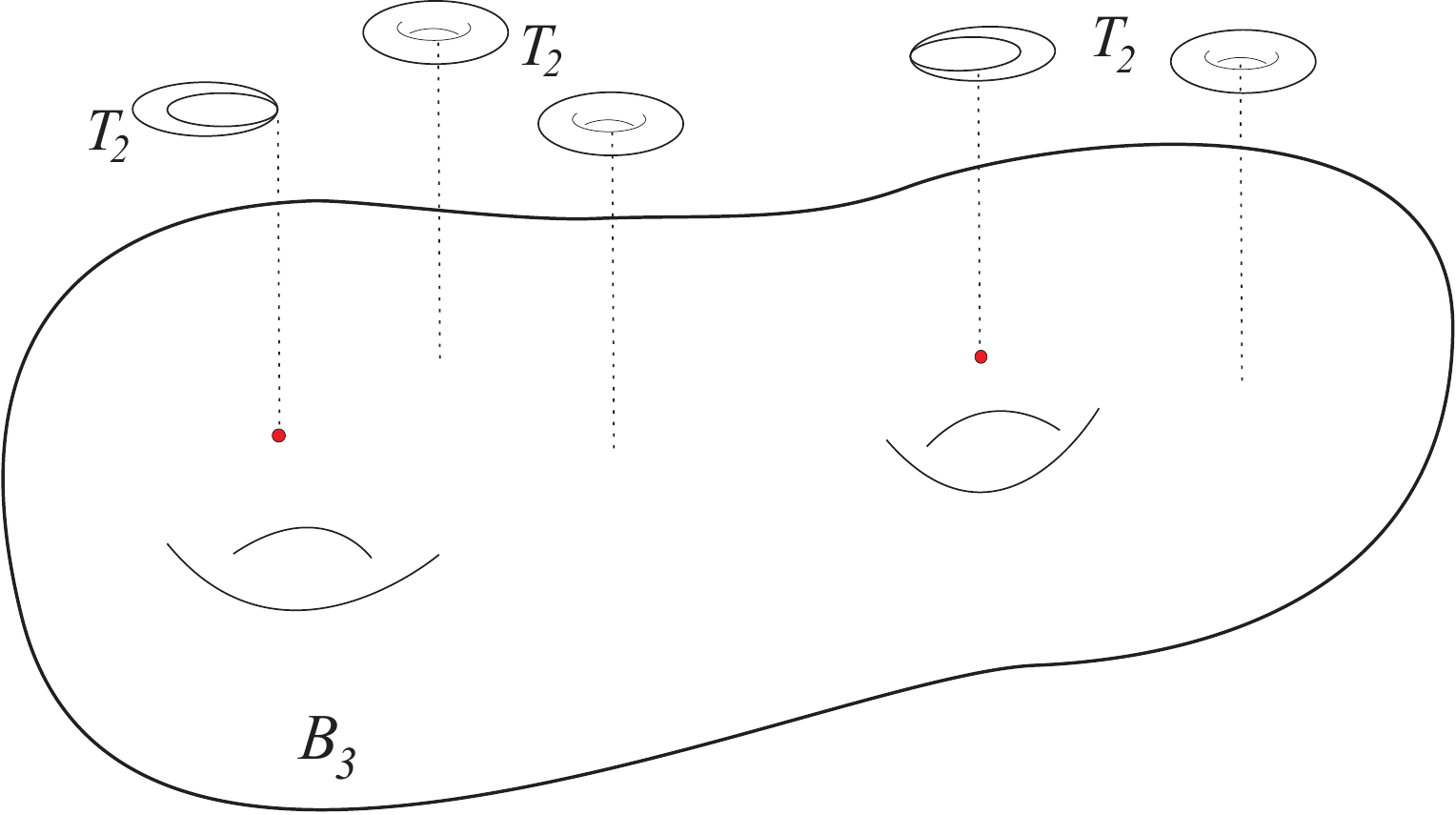}
\caption{Elliptic fibration. At each point of the threefold $B_3$ a torus  $\tau = C_0+ie^{-\phi}$ is assigned.}
\label{FiberB3}
\end{figure}

In the bosonic spectrum there are two scalars, the dilaton field $\phi$ and a zero-form potential
(axion) $C_0$. One then can define a  modulus,  the axion-dilaton complex structure
\be
 \tau = C_0+ie^{-\phi} \label{Cdilaton}
 \ee
and write down an $SL(2,Z)$ invariant action of the ten-dimensional theory
which leads to the correct equations of motion.  The terms of the action seem as if they are obtained
from a twelve dimensional theory compactified along the two radii of the torus (for a review
see~\cite{Denef:2008wq}).
We can think of $\tau$ as the modulus of a torus attached to each point of the  internal
manifold of three complex dimensions (threefold), as depicted in figure~\ref{FiberB3}.
We end up with a fibred  fourfold.
Recalling the analysis of the first section, one is tempted to  consider the interesting possibility
of describing  this fibration by the  Weierstra\ss\,   model given in equation (\ref{WeiF}).
In particular, we write it in the  form
\be y^2=x^3+f x z^4+g z^6\label{WeiF2}
\ee
where $x,y,z$ are homogeneous complex coordinates.
As explained previously, for $f,g$ complex,  equation   (\ref{WeiF}) describes a torus whose modulus
$\tau$ is now  identified with that of~(\ref{Cdilaton}).
On the other hand, in order to satisfy the Calabi-Yau  (CY) conditions, we also require the two
functions $f=f(w), g=g(w)$  to be $8^{th}$ and $12^{th}$ degree polynomials of the complex variable $w$.
As we move from point to point in the internal manifold, the modulus $\tau$ varies. In particular, on moving along
non-trivial closed cycles, $\tau$ undergoes non-trivial $SL(2,Z)$
transformations~\footnote{The $SL(2,Z)$  modular invariant function is given by
$j(\tau)= e^{-2\pi i\tau} +744+{\cal O}(e^{-2\pi i\tau})$. Combined with~(\ref{Jinv})
one can elaborate~\cite{Sen:1996vd} a relation approximated with $\tau(w)\sim \frac{1}{2\pi i}\ln(w-w_i)$
in the vicinity of the zeros of $\Delta(w_i)=0$.}.  In  figure~\ref{FiberB3}
for any generic point we draw a normal torus, while
pinched torii are drawn at points of singularities;
the latter appear when two D7-branes intersect at a `point' of the manifold.
These correspond to singularities  of elliptic surfaces and were
classified  in terms  of the vanishing orders of the discriminant and
the polynomials $f(w), g(w)$ several decades ago by Kodaira~\cite{Kodaira:1963}.
For minimal elliptic surfaces eight types of singular fiber were identified,
(nodal, cuspidal or otherwise reducible).
 The singularities are related to simply-laced Dynkin diagrams of ADE type.
These extremely interesting results can be found in several recent papers
and reviews~\cite{Kodaira:1963,Bershadsky:1996nh,Esole:2011sm}, thus, they will not
 be presented in this short note. Instead,  we will shortly give an analogous algorithm
 in another representation
 which will be useful in our subsequent analysis.

The nature of the singularities of the internal space  motivated the idea that
they can be identified with the gauge symmetries of the effective field theory
model.  If this is true, then one can attribute all the properties of the
internal manifold to the massless spectrum and  the effective potential
describing their interactions. This scenario has many  advantages, including
 calculability of Yukawa couplings~\cite{Heckman:2008qa}-\cite{Font:2013ida} of  the effective theory
 from a handful of geometric characteristics of the internal space.
\begin{table}[!t]
\centering
\renewcommand{\arraystretch}{1.2}
\begin{tabular}{|c|c|c|c|c|c|c|c|}
\hline
Type &{\bf Group} & ${ a_1 }$& ${ a_2 }$& ${ a_3} $& ${ a_4 }$& ${a_6} $& ${ \Delta}$\\
\hline
$I_0$& $-$&0&0&$0$&$0$&$0$&$0$   \\
\hline
$I_1$& $-$&0&0&$1$&$1$&$1$&$1$   \\
 \hline
$I_2$&  $SU(2)$&0&0&$1$&$1$&$2$&$2$   \\
\hline
$I^{ns}_{2m}$& $Sp(m)$&$0$&$0$&$m$&$m$&$2m$&$2m$\\
 \hline
$I_{2m}^s$& ${ SU(2m)}$&0&1&$m$&$m$&$2m$&$2m$   \\
\hline
$I_{2m+1}^s$& ${ SU(2m+1)}$&0&1&$m$&$m+1$&$2m+1$&$2m+1$   \\
 \hline
$I_1^{*s}$& ${ SO(10)}$&1&1&$2$&$3$&$5$&$7$   \\
 \hline
  $I_{2m-3}^{*ns}$& ${ SO(4m+1)}$&1&1&$m$&$m+1$&$2m$&$2m+3$   \\
   \hline
 $I_{2m-3}^{*s}$& ${ SO(4m+2)}$&1&1&$m$&$m+1$&$2m+1$&$2m+3$   \\
  \hline
  $I_{2m-2}^{*ns}$& ${ SO(4m+3)}$&1&1&$m+1$&$m+1$&$2m+1$&$2m+4$   \\
    \hline
      $I_{2m-2}^{*n}$& ${ SO(4m+4)^*}$&1&1&$m+1$&$m+1$&$2m+1$&$2m+4$   \\
        \hline
$IV^{*s}$& ${{\cal E}_6}$&1&2&$2$&$3$&$5$&$8$   \\
 \hline
$III^{*s}$& ${ {\cal E}_7}$&1&2&$3$&$3$&$5$&$9$   \\
 \hline
$II^{s}$& ${ {\cal E}_8}$&1&2&$3$&$4$&$5$&$10$  \\
 \hline
\end{tabular}
 \caption{Selected  cases of  Tate's algorithm.  The first column declares
 the type of the singular fiber according to Kodaira, i.e. nodal
 ($I_1$),  cuspidal ($II$) etc.
 The superscripts $s,ns$ stand for {\it split} and {\it non-split}.
 (The complete results can be found in~\cite{Tate1975,Bershadsky:1996nh,Esole:2011sm}.)
  The other columns show the order of vanishing of the coefficients
  ${ a_i\sim z^{n_i}}$,  the discriminant $\Delta$  and the corresponding gauge group.
  \label{TateTable}}
\end{table}

A convenient description which emphasizes the local properties
 of these singularities is given in terms of  Tate's algorithm~\cite{Tate1975}.
In this context, the equation   describing the elliptically fibred space takes the form
 \be
 y^2+\alpha_1 xyz + \alpha_3 yz^3
 = x^3+\alpha_2x^2z^2+\alpha_4 xz^4+\alpha_6z^6\label{TateAlg}
 \ee
The variables $[x,y,z]$ have weights $[2:3:1]$ correspondingly,
defining a hypersurface in the $\mathbf{P}_{(2,3,1)}$ weighted projective space.

In analogy with Kodaira's classification of singularities, here also the
gauge group is determined in terms of the vanishing orders of the
polynomials $\alpha_k$ and the discriminant $\Delta$. The results are summarised in Table
\ref{TateTable}.
 We note that the Weierstra\ss\, equation can be obtained from  Tate's form
by recovering the functions $f,g$ from the coefficients $\alpha_k$. To this end,
it is convenient to define the following quantities
\be
\beta_2=\alpha _1^2+4 \alpha _2;\;
\beta_4=\alpha _1 \alpha _3+2 \alpha _4;\;
\beta_6=\alpha _3^2+4 \alpha _6;\;
\beta_8=
\frac{1}{4} \left(\beta _2 \beta _6-\beta _4^2\right)\,.
\ee
Then, it can be readily checked that the functions $f,g$ and the discriminant $\Delta $ are
\ba
f&=&\frac{1}{48} \left(24 \beta _4-\beta _2^2\right)\\
g&=&\frac{1}{864} \left(\beta _2^3-36 \beta _4 \beta _2+216 \beta _6\right)\\
\Delta&=&-8 \beta _4^3+9 \beta _2 \beta _6 \beta _4-27 \beta _6^2-\beta _2^2 \beta _8
\ea
After these preliminary notes, in the next section we proceed to the description of the
basic tools for local model building.

\subsection{GUT models with discrete symmetries}

The attractive scenario of linking gauge symmetries to the singularities of the internal
geometry leads to far reaching implications.  An interesting advantage of  F-theory
constructions  based on the elliptic fibration, is the appearance of the exceptional  symmetry
${\cal E}_8$ where the  gauge group of the effective theory is
embedded\cite{Donagi:2008ca,Beasley:2008dc,Donagi:2008kj,Beasley:2008kw,Blumenhagen:2009yv}\footnote{For
reviews, see~\cite{Heckman:2010bq,Weigand:2010wm,Leontaris:2012mh,Maharana:2012tu}}. However,
phenomenological investigations have shown that additional symmetries (discrete or continuous)
are required to render the theory viable. Interestingly, a
 useful class of such symmetries originates from the commutant of this GUT
 with respect to  the exceptional gauge symmetry  ${\cal E}_8$.

To show how these symmetries appear we describe  the  ${\cal E}_6$  and $SU(5)$ gauge groups in brief.
In the local picture, Tate's coefficients  have a general expansion of the form
\be
\alpha_k=\alpha_{k0}+ \alpha_{k1}w+ \alpha_{k2}w^2+\dots\label{localexp1}
\ee
If  a certain coefficient $ \alpha_k$ has vanishing order $n$, it is convenient to write
\be
\alpha_k= \alpha_{k,n} w^n,\; {\rm with}\;  \alpha_{k,n} =  \alpha_{kn}+ \alpha_{k(n+1)}w+\cdots
\label{localexp2}
\ee
Hence, for an  ${\cal E}_6$ type of singularity, the coefficients take the form
\[ \alpha_1= \alpha _{1,1} w,\; \alpha_2 =\alpha _{2,2} w^2,\; \alpha_3=\alpha _{3,2} w^2,\;\alpha_4=\alpha _{4,3} w^3,\;
\alpha_6=\alpha _{6,5} w^5  \]
With this choice,  the discriminant is factorised as follows
\be
   \Delta =\Delta_0\, w^8\label{2lociD}
\ee
with
\be
 \Delta_0= -27 \alpha _{3,2}^4 +A(\alpha_{kj}) w +{\cal O}(w^{2})\\
\ee
where
\[
A(\alpha_{kj})  = \left(\alpha _{1,1} \alpha _{3,2}+2 \alpha _{4,3}\right)
     \left(\left(\alpha _{1,1}^2+36 \alpha _{2,2}\right) \alpha
     _{3,2}^2-32 \alpha _{4,3}( \alpha _{1,1} \alpha _{3,2}+ \alpha_{4,3})\right)
     -216 \alpha _{3,2}^2 \alpha _{6,5}\]
Indeed, $\Delta$ has a  vanishing order of $8^{th}$ degree, in accordance with Table~\ref{TateTable}.
From (\ref{2lociD}) we observe that the discriminant locus consists of two divisors,
 $ D_{{\cal E}_6}$ (at $ w=0$ of multiplicity eight) and   $ D_{{\cal I}}$ (at $\Delta_0=0$
 of multiplicity one).
There are eight $D7$ branes wrapping the divisor $ D_{{\cal E}_6}$ and one $D7$ brane wrapping
$\Delta_{{\cal I}}$ which is assumed to be irreducible.

The representations of the effective theory model, reside at the intersections of the
$ D_{{\cal E}_6}$ divisor  with D7 branes spanning different dimensions of the
internal space. These intersections (often called matter curves) are in fact Riemann surfaces along which symmetry
is enhanced. In the elliptic fibration the highest allowed singularity is ${\cal E}_8$.
Then, a convenient way to see the effective  model is  through the decomposition
\[  {\cal E}_8  \to {\cal E}_6\times SU(3)  \]
where ${\cal E}_6$ is the desired GUT, while  the enhancements along the matter curves include
factors embedded in $SU(3)$.   We can also think of $SU(3)$   broken by fluxes
(or some other mechanism) to a subgroup of it. The possibilities are either the
continuous symmetries
$SU(2), U(1)^2$, or  a discrete group such as the  $S_3$ (permutation of three objects),
$Z_3$ or $Z_2$. Hence  all  ${\cal E}_6$ representations transform non-trivially under the latter.
Viable cases have a final symmetry  such as~\cite{Beasley:2008kw,King:2010mq,Callaghan:2011jj,Callaghan:2012rv}:
\[ {\cal E}_6\times U(1)^2, \;  {\cal E}_6\times S_3,\; {\cal E}_6\times Z_2 \]

As a second example we consider that the GUT gauge symmetry is associated to a divisor characterised
by an $SU(5)$ singularity\footnote{F-$SU(5)$  models  have been extensively discussed in the
literature~\cite{Beasley:2008dc,Donagi:2008kj,Beasley:2008kw,Blumenhagen:2009yv,Dudas:2009hu,Grimm:2010ez,Braun:2011zm,King:2010mq,Callaghan:2011jj}},
 while  the commutant is also   $SU(5)$  -usually denoted with $SU(5)_{\perp}$.
It's a simple exercise to repeat the above analysis for the $SU(5)$ case too. Instead, let us focus on another
issue.    A phenomenologically friendly description of  these symmetries is based on the idea of the spectral cover.
In this case the  implications of $SU(5)_{\perp}$ are  described by a spectral cover denoted by ${\cal C}_5$
and represented by the five degree polynomial of an affine
coordinate $s$,
 \be
{\cal C}_5: \;\;\;\; \;\sum_{k=0}^5 b_ks^{5-k}\;=\; b_0s^5+b_1s^4+b_2s^3+b_3s^2+b_4s+b_5=0\label{sc5}
 \ee
The coefficients $b_k$ fulfill the conditions for the $SU(5)$ singularity ($ b_1=0$ for $SU(n)$)
\[  b_0=\alpha_{6,5},\; b_2=\alpha_{4,3},\; b_3=\alpha_{3,2},\;b_4=\alpha_{2,1},\; b_4=\alpha_{1,0} \]
 Equation (\ref{sc5}) includes the basic information regarding geometric properties as well as
additional symmetries of the $SU(5)$ F-GUT. Depending on the specific topological structure of the internal space,
 the spectral cover  ${\cal C}_5$ may factorise  in various ways. A few interesting cases are
 \[  {\cal C}_4\times  {\cal C}_1,\; {\cal C}_3\times  {\cal C}_2,\; {\cal C}_2\times  {\cal C}_2\times {\cal C}_1 \]
 implying analogous factorisations of the polynomial (\ref{sc5}). For the $SU(5)$ GUT, there is a rich variety of possible
accompanying discrete symmetries, including~\cite{Dudas:2009hu,Antoniadis:2013joa,Karozas:2014aha}
\[ SU(5)\times A_4\times U(1),\;SU(5)\times Z_3\times Z_2,\; SU(5)\times Z_2\times Z_2\times U(1)
 \]

\section{Mordell-Weil $U(1)$'s and discrete symmetries}

In the previous section we presented the classification of the non-abelian singularities
of the elliptic fiber,  subject to restrictions arising from Kodaira classification.
Adopting the interpretation that these correspond to non-abelian gauge symmetries, we were able
to determine the GUT gauge group of the potential F-theory models.
There is considerable  activity~\cite{Morrison:2012ei}-\cite{Esole:2014dea}  on the role of the abelian sector
related to the rational sections of the elliptic curves.  Subsequently, we focus in some related issues.

In the introductory section we have seen that there is a class of abelian symmetries,
associated to rational sections of the elliptic curves. Since the internal space
is elliptically fibred, these $U(1)$'s  may manifest themselves in potential
low energy effective models. From analyses of phenomenological models, we know
that such symmetries are extremely useful in order to prevent unwanted terms in the lagrangian.
It seems that such abelian  symmetries are indispensable when constructing an F-theory effective model,
however, a Kodaira-type classification is lacking up to now.
  From the Mordell-Weil theorem we only  know that  these are related to the rational
   sections defined on  elliptic curves but  the  rank of this group  is not known.
 The Mordell-Weil group can be written as
  \[{\cal E}({\cal K})= \underbrace{Z\oplus Z\oplus \cdots \oplus Z}_{n}\oplus {\cal G}\equiv  Z^n\oplus {\cal G} \]
 Here $n$ is the rank of the Abelian group, while ${\cal G}$ is the torsion subgroup and ${\cal K}$ is the number field.
 According to a theorem by Mazur~\cite{Mazur} (see also \cite{Kubert}), the possible torsion subgroups are either
$ \mathbf{Z}_k, k=1,2,\dots, 10,12$ or the direct  sum $ \mathbf{Z}_2\oplus  \mathbf{Z}_{2k}$ with $k=1,2,3,4$.
  \be  { \cal G}=  \left\{\begin{array}{ll}
                \mathbf{Z}_n &n=1,2, \dots, 10,12\\
                \mathbf{Z}_{2k}\oplus  \mathbf{Z}_2 &k=1,2,3,4\\
                 \end{array}
                 \right.
  \label{Tortion}
  \ee
  A specific choice of the coefficients in an elliptic curve equation
 eventually will fix the symmetries of the effective GUT model.
For a simple demonstration on the appearance of such symmetries, let us see how a $ \mathbf{Z}_2$
 discrete symmetry can arise~\cite{Aspinwall:1998xj}. Under a $ \mathbf{Z}_2$ action,  a  point $P$ on an elliptic curve
is identified with its opposite, $-P$. From the group law, (see figure~\ref{GroupLaw})
  $P+(-P) =O$, hence $P+P=O$, where for the Weierstra\ss\, form the zero element $O$ is taken to infinity.
 This implies that  line $OP$ must be tangent to $P$.  If we put $P$  at the origin, (0,0)
 then $dy/dx=\infty$. For example, the elliptic curve   $ y^2= x(x^2+x+1)$
 has a $ \mathbf{Z}_2$ symmetry at the origin (0,0).


\subsection{ On GUT Models with  Mordell-Weil  $U(1)$ 's}

In the geometric picture of  F-theory discussed previously, the elliptic fibration
assumed over a base $B_3$  can be defined as a holomorphic section of the fourfold.
In the following, the possibility of having a fibration of an
elliptic curve with two rational sections, including  the zero (universal) section will be examined.
 This  leads to a  rank-one Mordell-Weil group, or a theory with one  $U(1)$ symmetry in
 addition to the non-abelian GUT group.
 From the phenomenological point of view this is one of the most viable possibilities.
 GUT models with an additional abelian factor and perhaps a discrete symmetry
 arising from the torsion part (\ref{Tortion}), are probably
adequate to impose  sufficient constraints on superpotential terms.

 However, in general, it is not easy to identify the $U(1)$  symmetries by starting directly form the Weierstra\ss\, form.
 Instead, it is more feasible to start with a different   representation of the elliptic curve where the Mordell-Weil
  rank and other  characteristics are more transparent.  Once we have identified the abelian structure, in order to study the
  non-abelian group  we convert our equation to the ordinary Weierstra\ss\, form using the appropriate
   birational  transformation.
\begin{figure}[h]
\centering
\includegraphics[scale=.7]{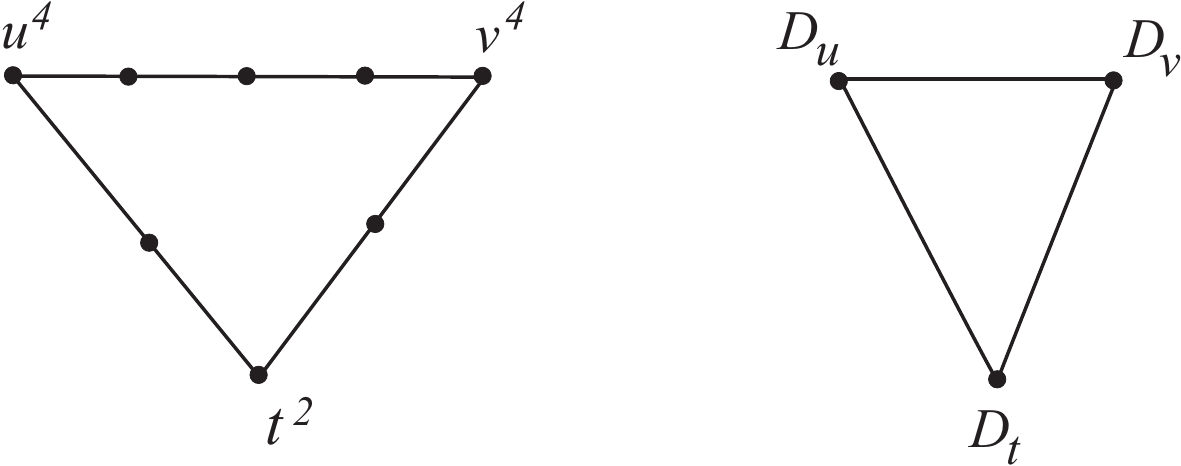}
\caption{The reflexible polygon for the model  $\mathbf{P}_{(1,1,2)}$  with one lattice point in its interior
in accordance with the genus-one elliptic curve. }
\label{ToricV}
\end{figure}
To derive the equation of such  a hypersurface,   following the analysis of~\cite{Morrison:2012ei},
we start with a point $P$ associated to the  holomorphic (zero) section and a rational  point $Q$ on an elliptic curve.
We introduce the degree-two line bundle  ${\cal M} = {\cal O}(P+Q)$ and denote  $u$ and $v$ its two independent
sections with weights $[1:1]$ generating the group $H^0({\cal M})$.
The space $H^0(2{\cal M})$ should have four independent sections. Given $u$ and $v$ we are able generate only three,
namely, $u^2,v^2$ and  $uv$. Thus we  need  to  introduce a new one, let $t$ with weight $2$, so we
are in a  $\mathbf{P}_{(1,1,2)}$  projective space of three sections $[u,v,t]$ with weights $[1:1:2]$ respectively.
 From $u,v,t$ we can form  six  sections  of degree 6 (namely $u^3,v^3,uv^2,u^2v$ and $tu,tv$)
 which match exactly the number of independent sections of
$H^0(3{\cal M})$. But  $u,v,t$ generate nine  sections for $H^0(4{\cal M})$
 exceeding the independent ones by one.  Hence  there has to be a constraint among them which defines
 a hyper-surface in the weighted projective space
 $\mathbf{P}_{(1,1,2)}$ given by  the equation of the form
\be
t^2+a_0u^2t+a_1uvt+a_2v^2t=b_0u^4+b_1u^3v+b_2u^2v^2+b_3uv^3+b_4 v^4
\label{hysu}
\ee
where $b_i, a_j$ are coefficients in the specific field ${\cal K}$ we are interested in.
The  $\mathbf{P}_{(1,1,2)}$ projective space can be regarded as a toric 
variety~\cite{Kreuzer:1995cd}-\cite{Grassi:2012qw} shown in
the left side of figure~\ref{ToricV}.  Furthermore, we can identify three divisors.
For $t=v=0, u\ne 0$  the divisor $D_u=[1:0:0]$, for $t=u=0, v\ne 0$  the divisor $D_v=[0:1:0]$,
and  for $u=v=0, t\ne 0$, the divisor $D_t=[0:0:1]$. These are indicated on the right
side of the same figure.
Without loss of generality~\cite{Morrison:2012ei} in order to avoid complications with
square roots etc,  we can simplify this equation to:
 \be
 t^2+a_2v^2t=u(b_0u^3+b_1u^2v+b_2uv^2+b_3v^3)
 \label{hysus}
 \ee
 Having constructed the elliptic curve equation with one Mordell-Weil $U(1)$, we would like now
 to transform this equation to the familiar  $\mathbf{P}_{(2,3,1)}$ model.
 In fact this is inevitable; in order to identify the non-abelian part of the gauge symmetry,
 we need to read off the singularity structure from the coefficients in the Weierstra\ss\, form.
 It can be proved that the conversion can occur by two sets of
 equations~\cite{Antoniadis:2014jma} relating the sections of the
 $\mathbf{P}_{(1,1,2)}$ model to those of  $\mathbf{P}_{(2,3,1)}$.  Both transformations
 lead to equivalent results. The simplest one is~\cite{Antoniadis:2014jma}:
\ba
v&=&\frac{a_2 y}{b_3^2 u^2-a_2^2 \left(b_2 u^2+x\right)}
  \\
  t&=&\frac{b_3 u y}{b_3^2 u^2-a_2^2 \left(b_2 u^2+x\right)}-\frac{x}{a_2}\\
    u&=&z
\ea
We substitute the above to $\mathbf{P}_{(1,1,2)}$ model and we recover the   Tate's form
\ba
{ y}^2+2\frac{b_3}{a_2}{ xyz }\pm b_1a_2{y}{ z}^3&=&{ x}^3\pm \left(b_2-\frac{b_3^2}{a_2^2}\right){ x}^2{ z}^2\nn\\
                                 &&-b_0a_2^2{ x}{ z}^4-b_0a_2^2\left(b_2-\frac{b_3^2}{a_2^2}\right){ z}^6
\nn
\ea
This is indeed in the desired $\mathbf{P}_{(2,3,1)}$ form  however not all of the
Tate's coefficients are independent. Comparing with the standard Tate's form
given in (\ref{TateAlg}) we observe that
\be
 \alpha_6    =  \alpha_2   \alpha_4  \label{A246}
\ee
As we shall see in the following, this relation inevitably implies  constraints  on the non-abelian singularities.
We restrict here the analysis in the Tate's form of the Weierstra\ss\, equation since it is
this form that we automatically obtain from the birational map.
Hence, we assume the local expansion of the Tate's coefficients which as a function of the ``normal''
coordinate they are given by (\ref{localexp1}) and (\ref{localexp2}).

To see the implications of the relation $\alpha_6=\alpha_4\alpha_2$, we need to substitute in it the specific
types of coefficients for each non-abelian group shown in Table~\ref{TateTable}.

We start the investigation with the $SU(n)$ singularities. According to Table~\ref{TateTable}
we must treat separately  even  $SU(2m)$ and odd  $SU(2m+1)$ cases.

\begin{enumerate}
\item
{\it $SU(2m)$ case.}
The vanishing orders of $\alpha_k$'s  for $SU(2n)$ groups are
 \[ \alpha_2 = \alpha_{2,1}w,\; \alpha_4 = \alpha_{4,m}w^m,\; \alpha_6 = \alpha_{6,2m}w^{2m}\]
 Substitution into (\ref{A246})  gives
 \[  \alpha_{2,1}\, \alpha_{4,m}w^{m+1} = \alpha_{6,2m}w^{2m}\]
 which is satisfied only for $m=1$, implying that only $SU(2)$ is compatible.

\item
{\it $SU(2m+1)$ case.}    Reading off the minimal powers of $\alpha_k$'s from
 Table~\ref{TateTable}, we get
 \[  \alpha_{2,1}\, \alpha_{4,m}w^{m+2} = \alpha_{6,2m}w^{2m+1}\]
This is also valid for $m=1$, hence only $SU(3)$ is admissible.
\end{enumerate}

Extending this analysis to the rest of the entries in Tate's table, one finds that the most interesting
cases arise for the exceptional groups.
We observe that under the particular birational map to Tate's form
the  only non-trivial admissible  non-abelian singularities   are   ${\cal E}_6$ and ${\cal E}_7$.

\begin{table}[!t]
\centering
\renewcommand{\arraystretch}{1.2}
\begin{tabular}{|c|c|c|c|c|c|c|c|}
\hline
Type &{\bf Group} & ${ a_1 }$& ${ a_2 }$& ${ a_3} $& ${ a_4 }$& ${a_6} $& ${ \Delta}$\\
\hline
$I_0$& $-$&0&0&$0$&$0$&$0$&$0$   \\
\hline
$I_1$& $-$&0&0&$1$&$1$&$1$&$1$   \\
 \hline
$I_{2}^s$& ${ SU(2)}$&0&1&$1$&$1$&$2$&$2$   \\
\hline
$I_{3}^s$& ${ SU(3)}$&0&1&$1$&$2$&$3$&$3$   \\
 \hline
  $I_{1}^{*ns}$& ${ SO(9)}$&1&1&$2$&$3$&$4$&$7$   \\
   \hline
      $I_{0}^{*n}$& ${ SO(8)^*}$&1&1&$2$&$2$&$3$&$6$   \\
        \hline
$IV^{*s}$& ${{\cal E}_6}$&1&2&$2$&$3$&$5$&$8$   \\
 \hline
$III^{*s}$& ${ {\cal E}_7}$&1&2&$3$&$3$&$5$&$9$   \\
 \hline
\end{tabular}
 \caption{Selected cases of Tate's  coefficients satisfying the relation
 $\alpha_6=\alpha_2\alpha_4$. The Standard Model is naturally embedded in the
 exceptional groups only.}
  \label{TateMordell}
\end{table}

\section{Conclusions}

In this talk, we have described a variety of discrete symmetries in F-theory models emerging in various ways.
Current F-theory constructions are based on the elliptically fibred internal space with ${\cal E}_8$
being the highest admissible singularity.
The non-abelian part of the gauge symmetry of a particular effective GUT model emerges as a
subgroup of the  ${\cal E}_8$.  However, in low energy effective models additional discrete or continuous
symmetries are required to  suppress  flavour changing processes and
prevent fast proton decay. To this end, two methods have been developed in the recent literature:

$\bullet$ A class of early F-theory  effective models are fully embedded in ${\cal E}_8$. As a consequence,
the GUT symmetry is a subgroup of ${\cal E}_8$ while the commutant incorporates any additional symmetry
of discrete or continuous nature which could be used to put constraints on the effective
lagrangian.

$\bullet$ In  recent works, abelian and discrete symmetries emerge
from the a non-trivial Mordell-Weil group, i.e. the group of rational points
of the elliptic curves associated to the fibration. In this approach, one
constructs a representation of the elliptic curve with the desired rational sections
and then finds  the birationally equivalent Weierstra\ss\,  equation.

In the presence of one Mordell-Weil factor in particular, we have shown that
the birational transformations to Tate's form gives two  viable gauge symmetries which  are
\[  {\cal E}_6 \times U(1)_{MW} \,\;{ \rm and} \,\; {\cal E}_7\times U(1)_{MW} \]
Although such a scenario looks rather restrictive, yet
these exceptional groups contain all the well known GUTs, such as
$SU(5)$, $SO(10)$ and the like, which can be readily obtained once we
break the initial symmetry  by a suitable mechanism, such as flux breaking, Wilson lines
mechanism etc.  Furthermore,  as long as the rank-one Mordell-Weil is
concerned, a novel generalisation of models in the context of elliptic fibrations
has been proposed recently~\cite{Esole:2014dea}.  Finally, it should be 
pointed out that discrete symmetries appear
naturally in F-theory compactifications without section
and examples with $SU(5)$ GUT symmetry
 have been presented in~\cite{Garcia-Etxebarria:2014qua}-\cite{Mayrhofer:2014haa}.

We close our discussion with a few observations. 
 The rather interesting fact in the procedure discussed in the last
 section, is that the $U(1)_{MW}$ symmetry
is not necessarily identified with some generator  of the Cartan  subalgebra
of ${\cal E}_8$. This means that the $U(1)_{MW}$ charges  of
the non-abelian representations are not necessarily the usual ones.
  Furthermore, the torsion group has a rich structure of discrete
 symmetries which can also be symmetries of the effective lagrangian.
 Perhaps issues such as proton stability, the $\mu$-term and flavour physics
 find their solutions in a judicious choice of these symmetries.

 {\ack} {\it  This research has been
co-financed by the European Union (European Social Fund - ESF) and
Greek national funds through the Operational Program ``Education and
Lifelong Learning'' of the National Strategic Reference Framework
(NSRF) - Research Funding Program: ``ARISTEIA''. Investing in the
society of knowledge through the European Social Fund.}

\end{document}